\documentclass[reprint,amssymb,amsmath,aip,cha]{revtex4-2}
\usepackage{graphicx}
\usepackage[caption = false]{subfig}
\usepackage{color}
\usepackage{epsfig}
\usepackage{ifpdf}
\usepackage{bm}
\usepackage[colorlinks=true,linkcolor=blue]{hyperref}%
\expandafter\ifx\csname package@font\endcsname\relax\else
 \expandafter\expandafter
 \expandafter\usepackage
 \expandafter\expandafter
 \expandafter{\csname package@font\endcsname}%
\fi
\usepackage{cleveref}
\begin{document}
\title{Plasma Agriculture: A green technology to attain the sustainable agriculture goal}
\author{Tanvira Malek} 
\affiliation{Institute of Advanced Research, Koba, Gandhinagar, 382426, Gujarat, India}
\author{Mangilal Choudhary}
\affiliation{Banasthali Vidiyapith, Tonk, 304022, Rajasthan, India}
\date{30-01-2023}
\begin{abstract}
The agriculture sector has many issues such as reductions of agricultural lands, growing population, health issues arising due to the use of synthetic fertilizers and pesticides, reduction in soil health due to extreme use of synthetic chemicals during farming, etc. The quality and quantity of foods required for living things are affected by many factors like scarcity of nutrient-rich soils, lack of suitable fertilizers, harmful insects and bugs, climate change, etc. There is a requirement to supply the proper nutrients to plants/crops for obtaining a high crop yield. Synthetic chemical fertilizers provide nutrients (macro and micro) to plants for their growth and development but the excess use of them is not good for a healthy lifestyle as well as for the environment. Plants need significant amounts of macro-nutrients (nitrogen, phosphorous, potassium, urea, etc.) and some micro-nutrients (iron, sulfur, magnesium, zinc, etc.) for growth and development through various physiological and metabolic processes of the plant system. Along with the nutrients, there is also a demand to control the harmful microbes, insects, pests, etc. during the growth of plants for increasing the crop yield. In recent years, non-thermal plasma (NTP) is considered as an advanced green technology for enhancing productivity in agriculture sectors. The plasma-treated water (PAW) can help in enhancing seeds germination, increasing the rooting speed, stimulating plant growth, deactivating microbes/bugs, etc. The atmospheric pressure plasma (NTP) contains energetic electrons, UV radiation, and various reactive nitrogen and oxygen species. During the plasma-water interaction, these reactive species in the gaseous form get dissolved into water and it becomes rich in nitrogen compounds (N-content). These nitrogen compounds in plasma-treated water act as fertilizer for plants to keep them healthy but PAW does not have some essential plant nutrients like potassium, phosphorus, sulfur, iron, magnesium, etc. Therefore, it is required to add such nutrients in addition to nitrogen compounds in the plasma-treated water to use it as a nutrient-rich fertilizer. In this report, we provided the details of nutrients and their functions in the growth and development of plants/crops. How plasma technology can resolve many future challenges in the agriculture sector is discussed in detail. A few experiments on seed germination and plant growth (root and shoot length) were performed in the laboratory to explore the effect of plasma-activated water on the growth and development of plants. These primary results demonstrate the great potential of plasma technology in the agriculture sector.   
\end{abstract}
\maketitle
\section{Introduction}
A continued increase in demand for food caused by exponential population growth indicates a serious challenge for humankind of the globe. All the time, plants/crops regularly face various stresses such as shortage of water, water-logging, toxicity, high saltiness, and excessive temperatures in some regions. There would be a significant effect of these stresses on crop yield. Another big issue for the agriculture sector is climate change day by day. Climate change causes a negative impact on the availability of food, reduction in access to food, quality of food due to polluted air and high temperature, etc. Same time we can also realize that agricultural land is continuously being reduced due to industrialization and urbanization. These all factors would be responsible for the shortage of food in the future \cite{agriculturechallenge1,agriculturechallenge2,agriculturechallenge5,agriculturechallenge4,agriculturechallenge3}. There is demand for improving the sustainability of agriculture and at the same time need to reduce the adverse effects of agriculture on the environment. To achieve these goals, new Eco-friendly technologies that can enhance productivity while maintaining food quality and safety are required. With the help of these green technologies in agriculture, it is possible to increase crop yields by enhancing productivity without damaging the environment and compromising human health. Crop productivity can be increased by keeping crops/plants healthy and providing the required nutrients along with water for growth and development. The use of high-quality seeds, fertilizer, pesticides, insecticides, suitable soil for plants/crops, etc. is the major deciding factor for the growth and development of healthy crops/plants. There is also a need to use good-quality seeds for higher crop yields \cite{agriculturechallenge4,agriculturechallenge1}. In recent years, researchers are working on new technologies to modify the seed morphology, increase the protein level, deactivate the seeds microbes, etc. for improving the seed germination rate and healthy and vigorous growth of plants/crops \cite{newagriculturetechnology2,newagriculturetech3,newagriculturetechnology1}. Along with some new technologies in agriculture, low-temperature or non-thermal plasma technology has been a popular green technology to use in the agriculture sector. The non-thermal plasma technology (NTP) has received considerable attention in recent years due to its increasing applications in the treatment of seeds and plants for enhancing germination rate and growth rates \cite{plasmaagriculturereview1,plasmaagriculturereview2}. The objective of this project was to identify the role of different macro and micro-nutrients in the growth and development of plants/crops and review the plasma technology with challenges in adopting by farmers.    
In this report, we have discussed the required nutrients for the growth and development of healthy plants/crops in Sec.~\ref{sec:secI}. Introduction to plasma and its interaction with water is discussed in Sec.~\ref{sec:secII}. Could we use plasma technology in farming? The answer to this question is given in Sec.~\ref{sec:secIII}. The challenges in the implementation of plasma technologies from lab to field are discussed in Sec.~\ref{sec:secIV}. The factors affecting seeds germination and the selection of seeds for conducting the experimental study are presented in Sec.~\ref{sec:secV}. Experimental setup and methods are discussed in Sec.~\ref{sec:secVI}. Primary experimental findings on seed germination and plant growth are presented in Sec.~\ref{sec:secVII}. Concluding remarks along with future perspectives on plasma agriculture are given in Sec.~\ref{sec:secVIII}
\section{Nutrients Requirement for plants} \label{sec:secI}
It is well known that in scaling up the plasma technology from the lab to the farm, the plant/crop physiology needs to be reviewed in detail. There are many processes in the plants such as transportation of minerals and nutrition, photosynthesis, respiration etc. these help for the overall growth and development of plants. 
Photosynthesis is essential to produce food (ATP and NADPH) for plants. In this process, there is a fixation of $CO_2$ in presence of incident sunlight. The intensity of incident light, carbon dioxide concentration, environment temperature, and  water are major affecting factors to modify the photosynthesis rates in green plants/crops. The food synthesised by the leaves, minerals and nutrients from roots has to be moved to all parts of the plant. There are various transportation processes such as diffusion, facilitated diffusion, transpiration stream and active transport in plants to transport water, mineral salts, some organic nitrogen and hormones from roots to the aerial parts of the plants and synthesised food from leaves to other parts of plants. It is fact that all plants/crops need some absolutely essential nutrients for growth and development. These elements are divided into two broad categories based on their quantitative requirements for plants/crops. (I) Macro-nutrients and (II) Micro-nutrients. The macro-nutrients include carbon, hydrogen, oxygen, nitrogen, phosphorous, sulphur, potassium, calcium and magnesium. And Micro-nutrients include iron, manganese, copper, molybdenum, zinc, boron, chlorine and nickel. Apart from carbon, hydrogen and oxygen, nitrogen is the most prevalent element in living organisms. Nitrogen is a constituent of amino acids, proteins, hormones, chlorophyll and many vitamins. Nitrogen is absorbed by roots form of $NO{_3}^{-}$ and $NH_4^{+}$ and transported to all parts of the plant for growth and development. Absorption of $NO{_3}^{-}$ and $NH_4^{+}$ are mainly affected by mainly concentration of these ions, temperature, pH of soil etc. phosphorus, potassium, sulphur, calcium, magnesium, zinc, iron, copper etc. all these nutrients are absorbed by the plants from the soil in the form of their ions. All these nutrients are involved in different reactions which are essential for the growth and development of healthy plants \cite{plantncert,plantnutrient1,plantgrowth1,plantgrowth2}. For example, phosphorus is required for phosphorylation reactions, potassium helps to maintain an anion-cation balance in cells and is involved in protein synthesis, calcium is involved in the functioning of the cell membrane and activates certain enzymes to regulate metabolic activities, sulphur is the main constituent of several enzymes and vitamins, magnesium activates the enzymes of respiration, photosynthesis and maintains the ribosome structure, iron is essential for the formation of chlorophyll, chlorine is essential for the water-splitting reaction in photosynthesis. It can be concluded that incomplete macro and micro-nutrients can lead to obstacles to the growth and development of plants and result in low crop yield at higher input costs \cite{plantncert,plantnutrient1,plantbook2}.
\begin{figure*} 
 \centering
\subfloat{{\includegraphics[scale=0.45050]{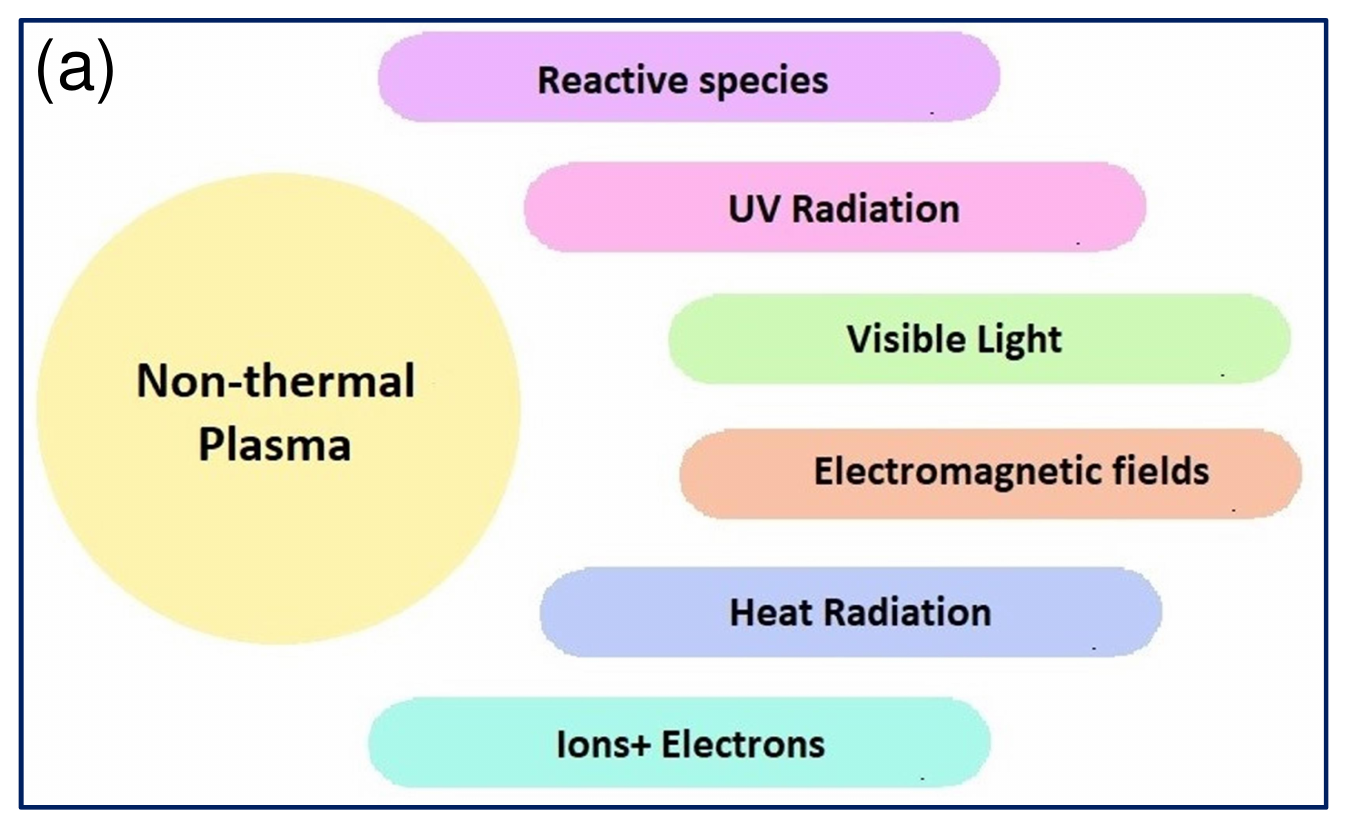}}}%
\hspace*{0.1in}
 \subfloat{{\includegraphics[scale=0.45050]{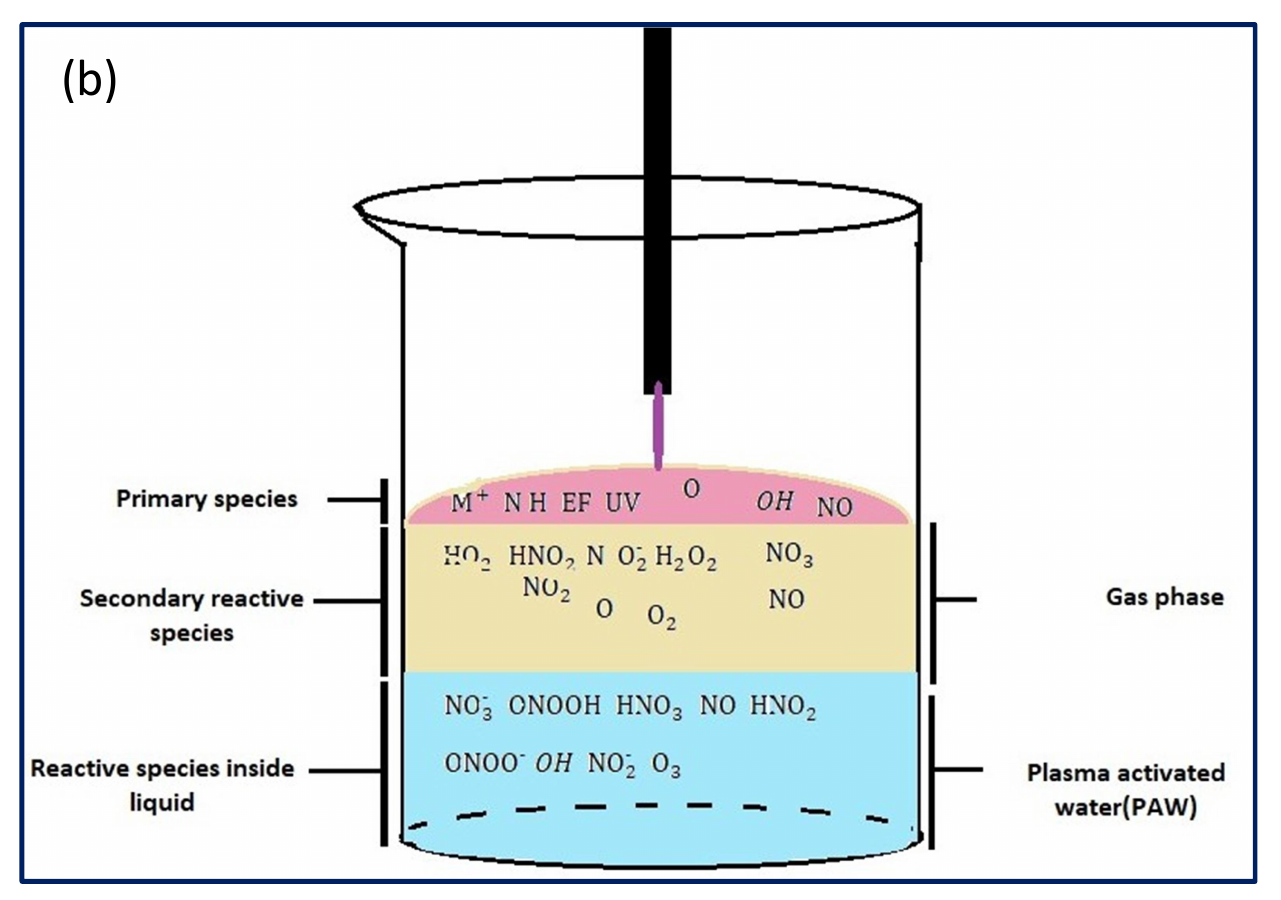}}}
 \quad
\caption{\label{fig:fig1}(a) A representation of non-thermal plasma and its constituents, (b) Non-thermal air plasma after interaction with water} 
\end{figure*}
\section{Non-thermal plasma and its interaction with water} \label{sec:secII}
As we know that plasma is one of the four common states of matter. It is an electrically charged gas consisting of charged particles (electrons and ions) that are not free but the motion of these charged particles is affected by electrical and magnetic fields of other moving charges. Plasma is created when the gas atoms are ionized by supplying external energy (electric energy). Based on the average energy of electrons and ions (neutrals), plasma can be characterized as thermal plasma and non-thermal plasma. In thermal plasma, the energy of electrons and ions (neutrals) are very large and nearly equal ($T_e$ = $T_i$ or $T_n$) whereas the average energy of electrons is higher than the energy of ions in non-thermal plasma ($T_e>>T_i$). These non-thermal plasma are in a non-equilibrium state and therefore have many advantages to applying in various sectors \cite{liebermanbook,ntpbook1}. The non-thermal plasma (air or $N_2/O_2$) plasma either in the gaseous phase or plasma-treated water has a great potential to contribute to the agriculture and food industries. Non-thermal plasma discharge, as shown in Fig.~\ref{fig:fig1} (a), is a source of visible and UV radiations, energetic electrons, excited atoms and molecules, various reactive oxygen and nitrogen species (RONS), various radicals, etc \cite{atmosphericplasma1,atmosphericplasma2,atmosphericplasma3,ntpbook1,ntpreview2}. \\
If non-thermal plasma (NTP) interacts with water or liquid solution then reactive oxygen and nitrogen species (RONS), energetic electrons, and radiations generated by NTP or atmospheric pressure plasma in the gaseous phase are transported through the plasma–liquid interface into the water (solution). The water or water solution after interaction with plasma is termed "plasma-activated water (PAW) or solution". The plasma-activated water has a different chemical composition such as superoxide ($O_2^{- .}$), hydroxyl radical ($OH^.$), oxides of nitrogen ($NO_2^{-}, NO_3^{-}, NO_2^{.}, NO_2^{.}, NO$), Hydrogen peroxides ($H_2O_2$), Ozone ($O_3$), singlet oxygen ($O^.$), Hypochlorous acid ($HOCl$), etc. than untreated water or simple water solution. A schematic image of non-thermal plasma to represent the reactive species before and after the plasma-water interaction is shown in Fig.~\ref{fig:fig1} (b). In other words, we can say that plasma-treated water contains significant amounts of reactive oxygen and nitrogen species \cite{paw1,paw2,paw3,pawreactivespecices1}

\section{Non-thermal plasma as an alternative in agriculture sector} \label{sec:secIII}
It has been discussed in the previous section that in presence of sunlight, plants produce foods using carbon dioxide and water. It is fact that maximum photosynthesis takes place in the red and blue light of the visible spectrum and minimum photosynthesis takes place in the green light. Dielectric barrier discharge (DBD) is one of the popular non-thermal plasma sources which can be used to generate a visible spectrum of radiations using a mixture of suitable gases to promote the photosynthesis process in plants/crops. In other words, the non-thermal plasma source can be used as a source of the visible spectrum that is essential for the photosynthesis reactions in plants \cite{atmosphericplasma1,atmosphericplasma2,atmosphericplasma3}. Apart from the visible spectrum of light, UV radiation also plays a major role in the growth of plants such as UV-A and UV-B (wavelength 315-400 nm and 280-315 nm respectively) are responsible for healthy and vigorous growth of plants but excess amount of UV-C (100-280 nm) can decrease the photosynthesis process. Non-thermal plasma in the gaseous phase contains a spectrum of UV radiation that can be useful for the growth and development of plants/crops. Thus, atmospheric pressure plasma (mixture of gases) can be used as an artificial source of sunlight to initiate the photosynthesis process which is essential for the growth and development of plants/crops. As we discussed that plasma treated water or solution has many RONS (radicals and non-radicals) which work as fertilizers, pesticides, and sources of macro-nutrients for the better growth of plants. The plasma-activated water is rich in nitrogen content ($NO_2^{-}, NO_3^{-}, NH_4^+$) that can be a promising alternative organic fertilizer to conventional chemical nitrogen fertilizers. The plasma-treated water or water solution shows antibacterial and fungicidal properties because of the presence of ozone ($O_3$) which works as a disinfectant and hydrogen peroxide ($H_2O_2$) which works as a pesticide \cite{pawreactivespecices1,plasmainactivation2,plasmaagriculture1,plasmaagriculturereview2,plasmainactivation3,plasmainactivation3,plasmainactivation4,plasmainactivation6}. In summary, the non-thermal plasma or atmospheric pressure plasma has great potential to replace conventional chemical fertilizers/pesticides and sources of light in futuristic agricultural developments.
\section{Non-thermal plasma technology and challenges} \label{sec:secIV}  
After a literature survey of plant physiology and non-thermal plasmas (or atmospheric pressure plasma, we found that it is possible to fulfill all the requirements of the plants/crops to grow just using non-thermal plasma as a source of UV radiation, visible light, macro-nutrients, pesticides, fertilizers, etc. In recent years, many research groups around the globe have started working on the application of low-temperature plasma in treating seeds, increasing the germination rate of seeds \cite{seedgermination2,seedgermination3,seedgermination3,seedgermination4,seedgermination5,seedgerminationwater6,seedsgermination1}, enhancing the growth of plants/crops \cite{plasmaplantgrowth1,plasmaplant3,plasmaplantros2}, deactivating microbes on fruits/vegetables \cite{plasmaindeactivation1,plasmainactivation2,plasmainactivation3,plasmainactivation4,plasmainactivation6}, and treating agriculture soils \cite{plasmasoil2,plasmasoil1}. There is great potential in plasma technology to improve crop yields by implementing it at various stages of the plant/crop life cycle But there are many challenges to scaling up the plasma technology from the lab to the field\cite{plasmareview2,plasmaagriculturereview1,plasmaagriculture1,plasmaagriculturereview2}. As we discussed that plasma-activated water (rich N-content) can be used as liquid fertilizer in place of nitrogenous chemical fertilizer \cite{paw1,pawreactivespecices1,plasmaagriculture1,plasmaagriculturereview1,plasmaagrirevpaw} but the same time plants/crops need other macro-nutrients (potassium, phosphorus, sulfur, etc.) and micro-nutrients (magnesium, zinc, iron, calcium, etc.). The plasma-activated water does not have these macro- and micro-nutrients that are essential for the growth and development of plants. Therefore, we need to add these macro- and micro-nutrients in the plasma-activated water to make it a complete liquid fertilizer. The second issue is the high energy cost to treat the water with plasma sources. Farmers may or may not use the costly liquid fertilizer in place of chemical fertilizer until the cost of liquid fertilizer would be reduced. We have started working on resolving some of these issues and will discuss them in an upcoming research article in detail. Using renewable energy sources such as solar cells, wind turbines, etc. to operate the plasma reactors for the treatment of water could be a solution for the high energy cost. The specific design of plasma reactors/use of appropriate discharges to treat the water or water solution can also be beneficial to reduce the energy cost of plasma reactors \cite{plasmaagrirevpaw}. 
\section{Seeds selection and germination affecting factors} \label{sec:secV}
In the present work, our focus was to study the seed germination rate with and without plasma-treated tap water. There were many open questions in mind before starting the experiments such as the selection of seeds, normal germination days, seeds anatomy, etc. Before going to the selection of seed we must have basic knowledge about plant family. Knowledge of plant families can help us to know about the germination of seeds from a particular family. If we know the characteristics of any one plant/crop of a particular family after plasma treatment then it is easy to get the characteristics of the whole family. As per season (May-June), we found that the legume family (moong, clovers, cowpeas, pulses, groundnut, etc.) and Gourd family (cucumber, pumpkin, melons, bottle guard, watermelons, etc.) are suitable for the experiment because of its capacity to grow in the summer season. We have selected the moong (vigna radiata) crop from the gourd family for the experiments due to its less germination time. Water uptake is essential for seed germination. Apart from the uptake of water, moisture, temperature, oxygen, and light are the main affecting factors of seed germination \cite{plantncert,plantbook2} We have performed all experiments by keeping all these information about seed germination in mind. 
\begin{figure*} 
 \centering
\subfloat{{\includegraphics[scale=0.4500050]{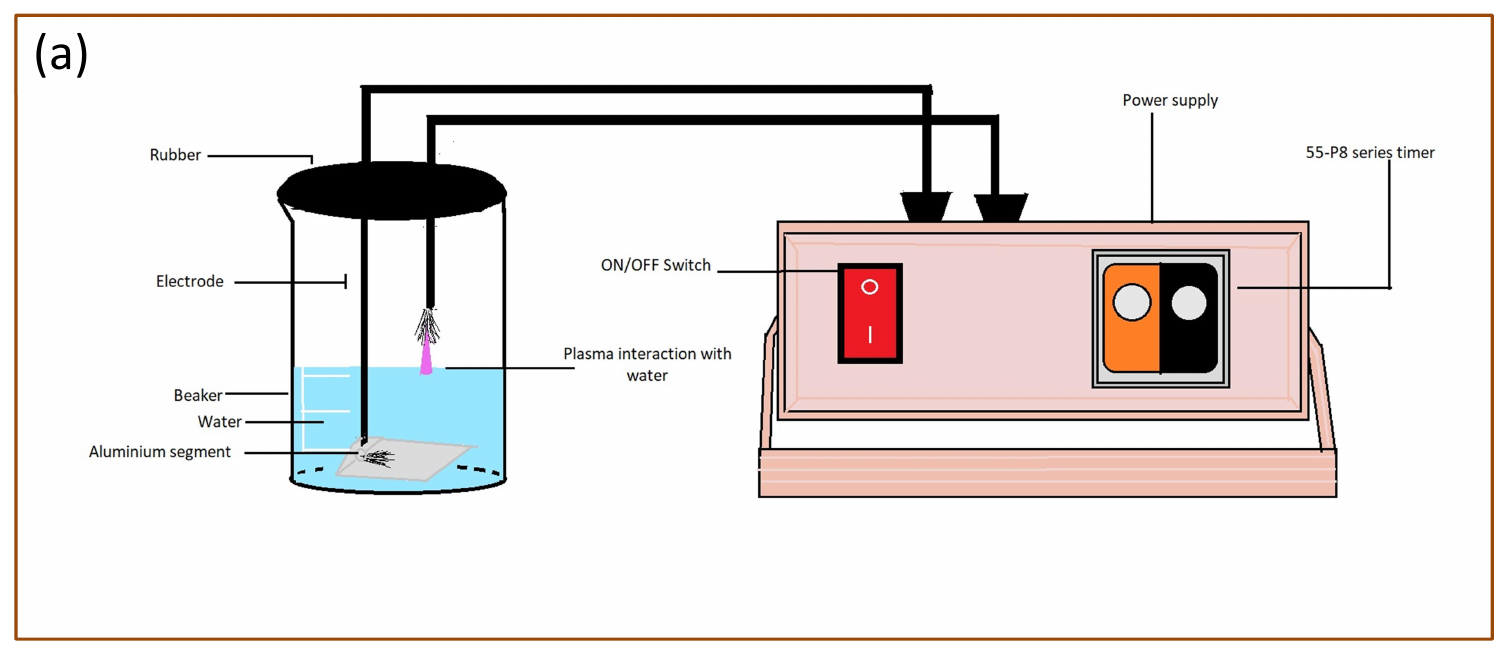}}}%
\hspace*{0.05in}
 \subfloat{{\includegraphics[scale=0.450050]{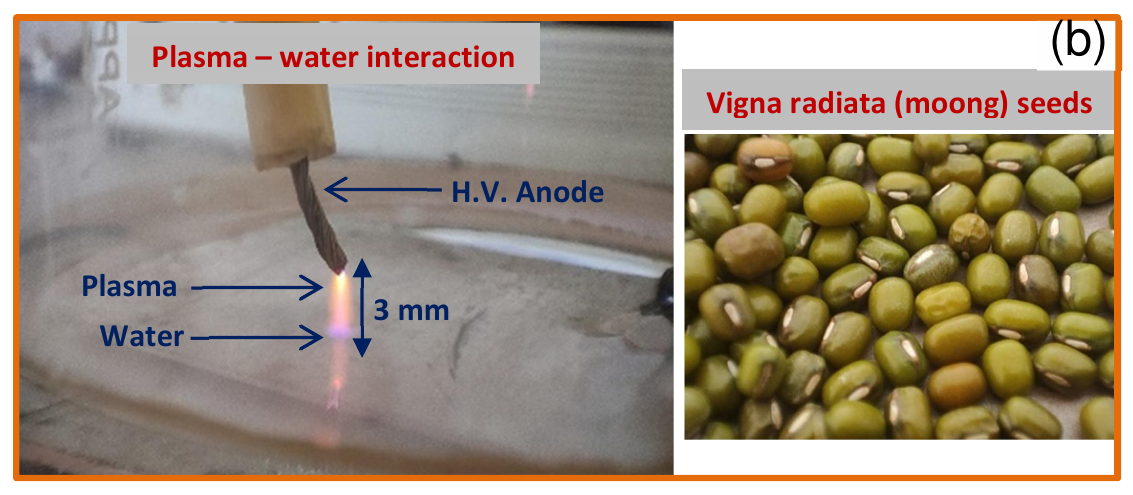}}}
\caption{\label{fig:fig2} (a) A Schematic diagram of the experimental setup (b) Image of plasma-water interaction and seeds used in the experimental study} 
\end{figure*}
\section{Experimental setup and methods} \label{sec:secVI}
\begin{figure*} 
 \centering
\subfloat{{\includegraphics[scale=0.30050]{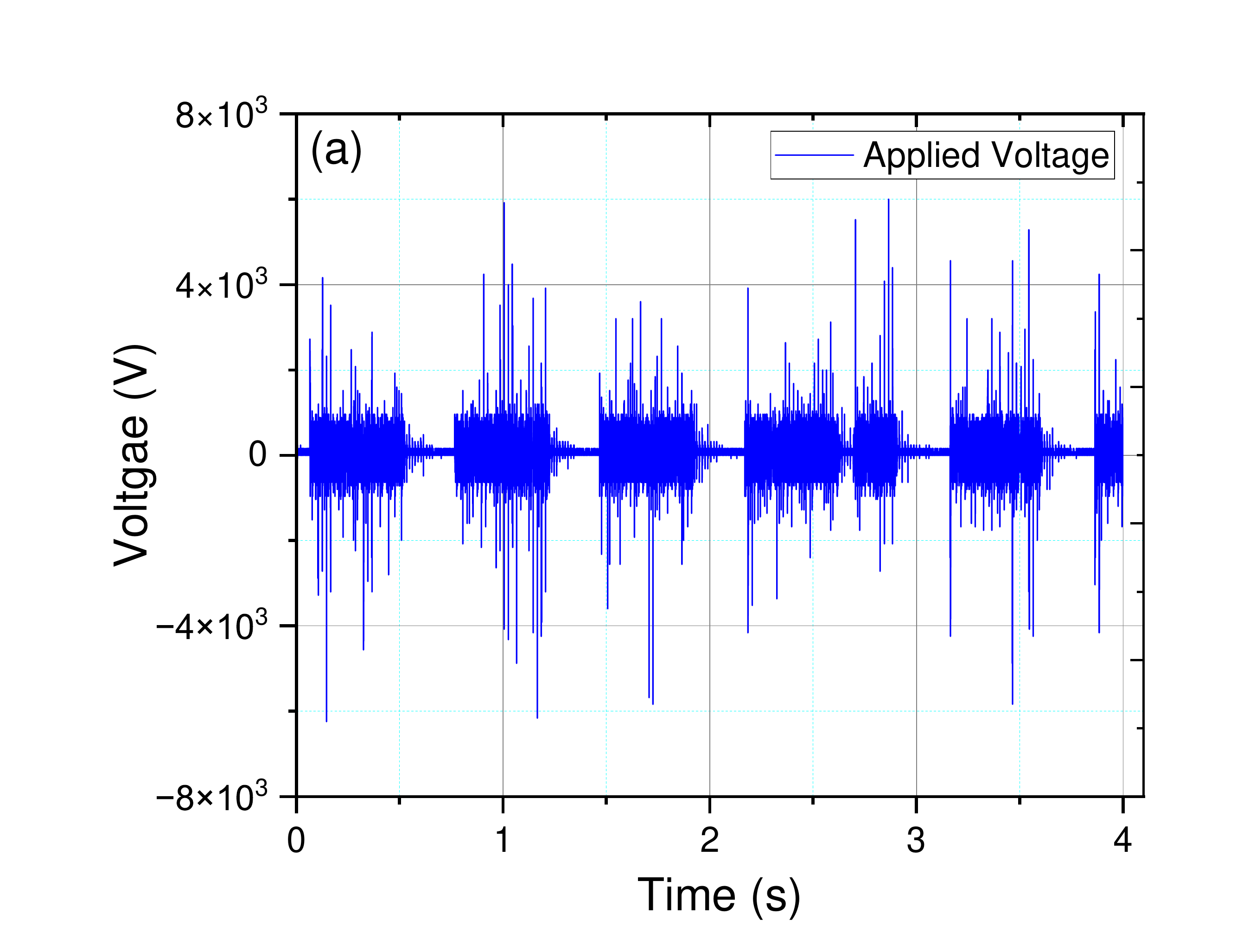}}}%
 \subfloat{{\includegraphics[scale=0.30050]{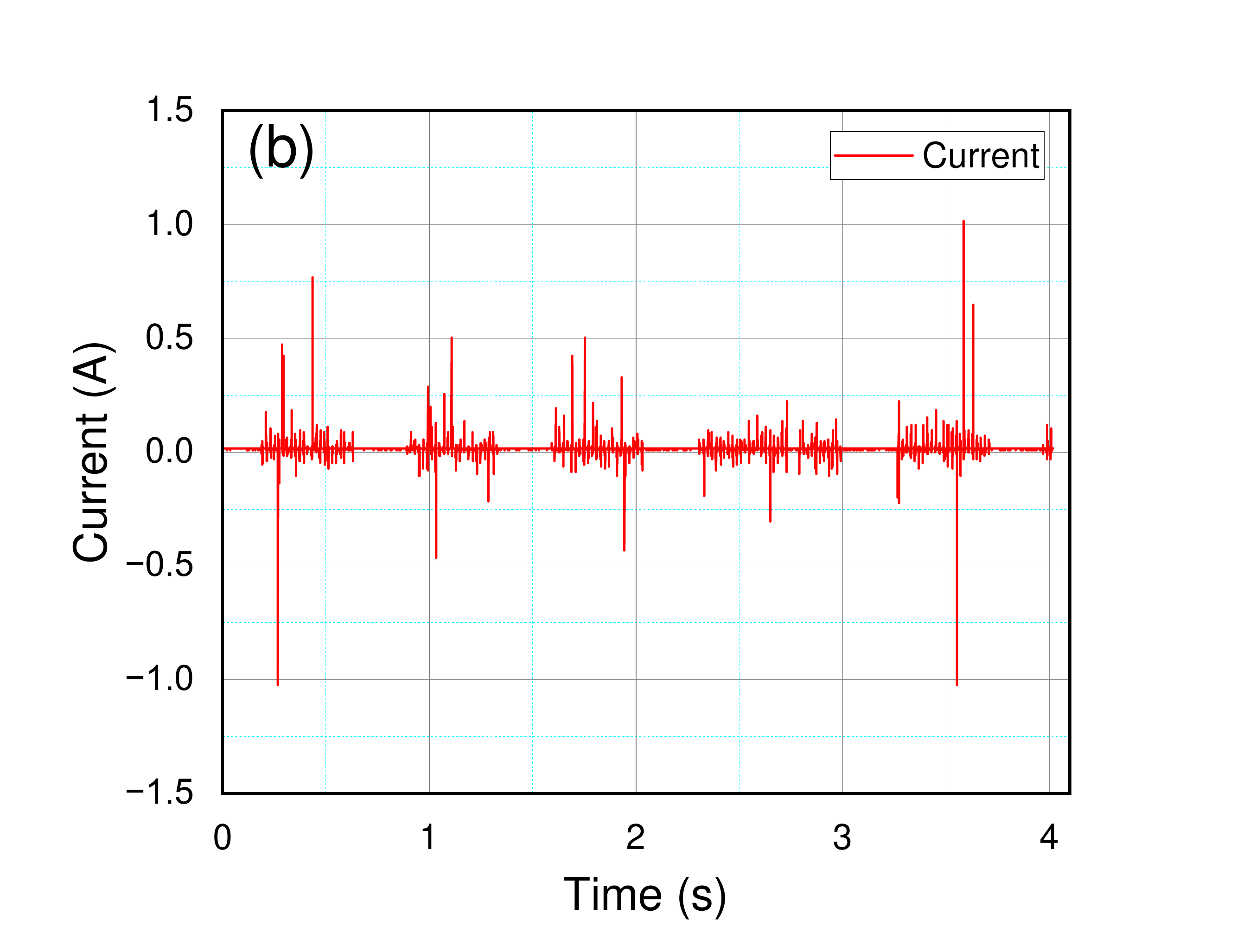}}}
\caption{\label{fig:fig3}(a) Applied transient high voltage pulses (b) Plasma current pulses} 
\end{figure*}

In the present study, we used a commercially available high voltage ($V_{p-p}$ = 6 kV) and low current ($<$ 1 A) power supply to treat the tap water. A schematic diagram of the experimental setup is shown in Fig.~\ref{fig:fig2}(a). There was a provision on the power supply to set the on and off time of sparking between the high voltage (H.V.) electrode and the grounded cathode. The high melting point alloy wire of diameter 3 mm as H.V. electrode (anode) and rectangular shaped (40 mm $\times$ 20 mm) grounded electrode made of aluminum was used (see Fig.~\ref{fig:fig2}(b)) as a cathode. A high-voltage probe and a coil loop (typical current transformer) were used to observe the applied voltage and corresponding current profile during the discharge. The applied voltage and corresponding plasma current burst (pulses) are shown in Fig.~\ref{fig:fig3}(a) and Fig.~\ref{fig:fig3}(b) respectively.  
 A Thermometer was used to measure the temperature and a pH meter for measuring the acidity or basicity (pH value) of the plasma-treated water at different plasma treatment times. The good quality moong seeds (vigna radiata) were purchased from the market. Other equipment and accessories such as beakers, pipettes, petri dishes, weight machines, stopwatch, etc. were used for conducting the primary experiments on seed germination with plasma-treated water.\\
\begin{figure} 
\centering
 \includegraphics[scale= 0.5500000]{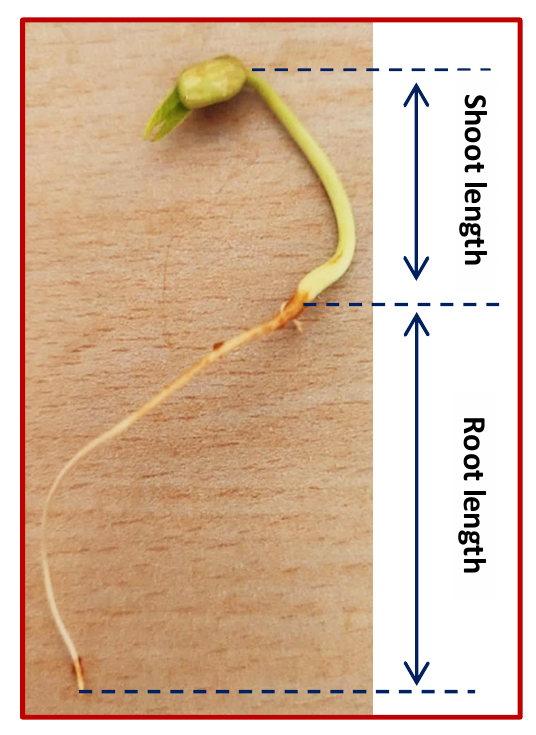}
\caption{\label{fig:fig4} Measurement technique for root and shoot length of a plant}
\end{figure} 
The effect of plasma-activated water on the seed's germination is verified by measuring the various parameters such as the pH of plasma-activated solutions, the temperature of the water, seeds germination ratio, root and shoot length, etc. The seeds germination ratio is germinated seeds divided by the total seeds taken in a sample. The root and shoot length were measured by taking measurements as shown in Fig.~\ref{fig:fig4} for some germinated seeds. The following steps were taken in performing the experiment and measurements:    
\begin{itemize}
\item We take an appropriate volume of tap water in the glass beaker
\item Two electrodes (cathode and anode) of power supply soaked into a beaker (200 ml) using an insulator feed-through. Cathode (grounded) is dipped into the water while the high voltage electrode (anode) is kept floating in the beaker 2 to 3 mm above the liquid surface. 
\item Turn on the H.V. power supply before setting the on and off time using the timer knob for water treatment.  
\item The non-thermal air transient spark discharge plasma is formed between the H.V. electrode and the water surface and it gets diffused into the water. The water is treated by plasma-water interaction and we get plasma-activated water.
\item  After the desirable plasma treatment, we turn off the power supply and use the plasma-activated or treated solution for further application.
\item First we measure the pH of plasma-treated water at different times and then use it to study the seed's germination.
\item We put the plasma-treated water (25 ml) into a Petri dish and add 20 seeds of Vigna radiata (moong).
\item We track the germination of seeds for different time intervals (hours or days)
\item  Measure the seeds germination coefficient, root length, and shoot length of growing plants at different time intervals (hours or days)
\end{itemize}
\section{Experimental results on seeds germination} \label{sec:secVII}
\begin{figure*} 
\centering
 \includegraphics[scale= 0.6500000]{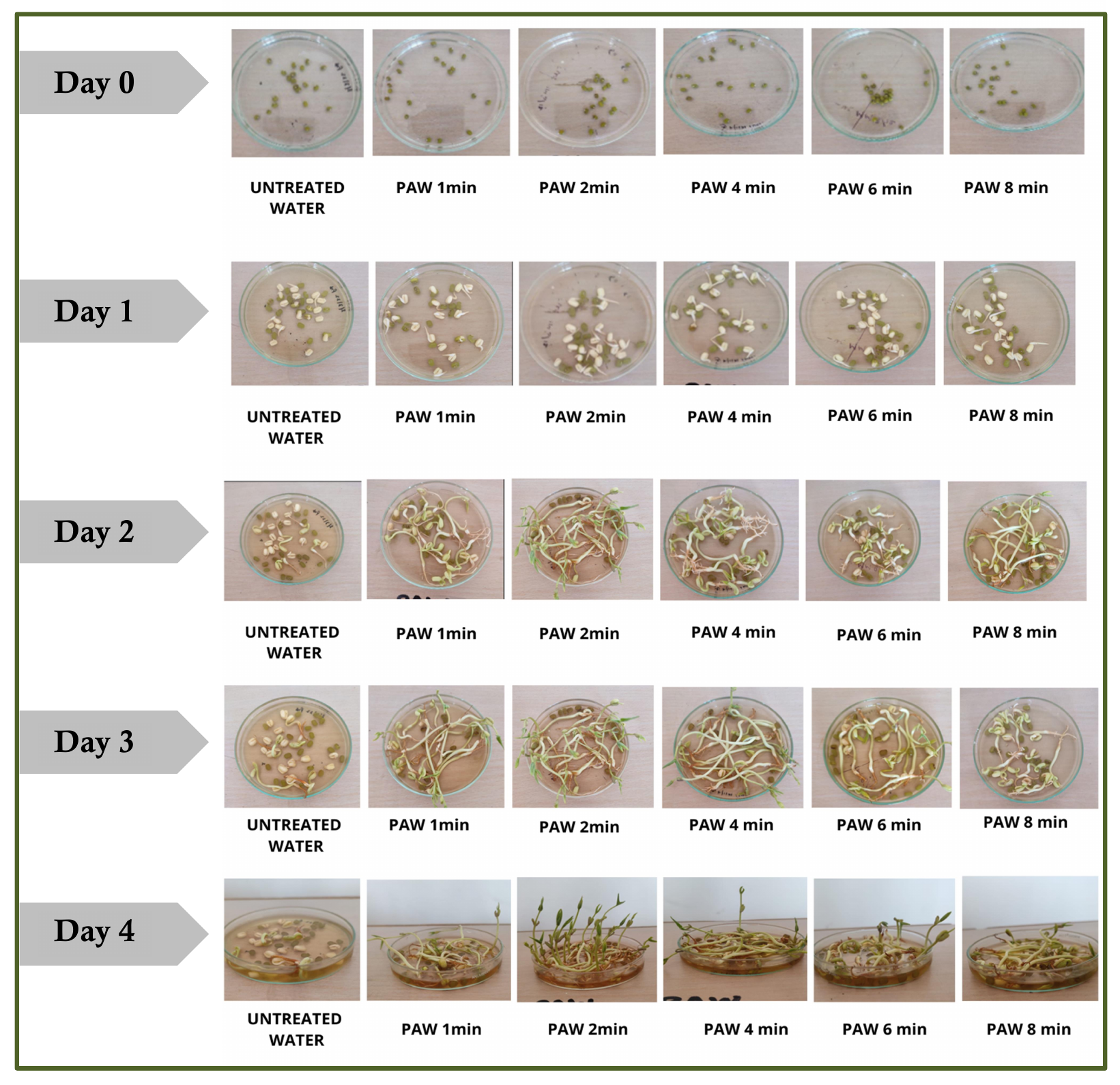}
\caption{\label{fig:fig5} Images of growth of Vigna radiata (moong seeds) in plasma-activated water samples at different times}
\end{figure*}

\begin{figure} 
\centering
 \includegraphics[scale= 0.3000000]{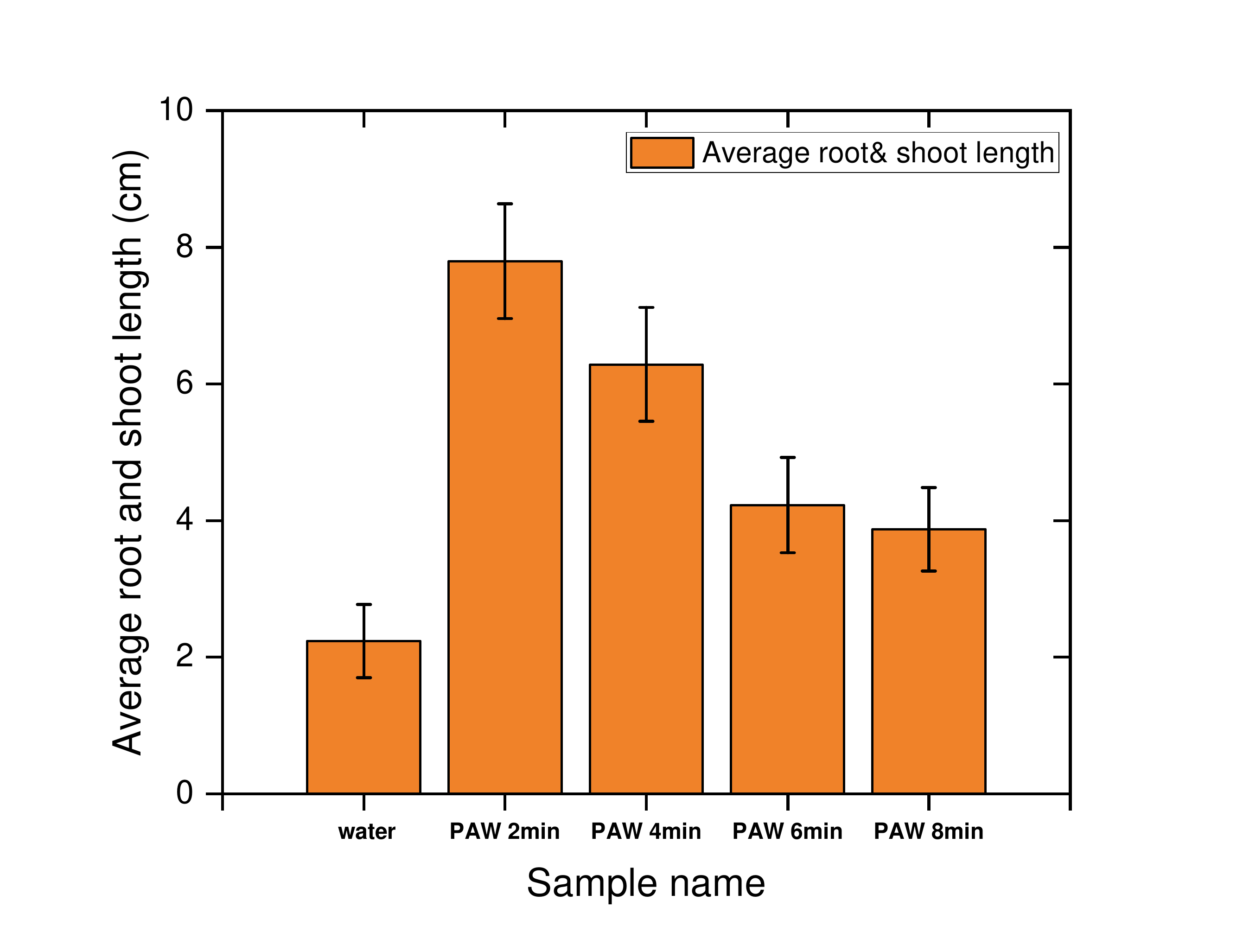}
\caption{\label{fig:fig6} Variation of root and shoot length (both) on the third day (after 72 hours) of soaking seeds in different plasma-activated water samples}
\end{figure}

In the first set of experiments, 25 ml of tap water was treated with atmospheric pressure air plasma at different times. We prepared 5 samples of plasma-treated water (25 ml each) based on the treatment time. The water (25 ml) treated by plasma for 1 min is named PAW 1min. Similarly, PAW 2min, PAW 4min, PAW 6min, and PAW 8min were prepared and poured into different Petri dishes. Then added 20 seeds of Vigna radiata (moong) into each plasma-activated water-containing Petri dish. The effect of plasma-activated water on seed germination and plant growth (root and shoot length) on different days is shown in Fig.~\ref{fig:fig5}. The seed's germination rate was tracked after 24 hours for four to five days. Seeds germination data on different days are given in Table-I. plasma-activated water for 2 min and normal water has more than 95 \% germination but plasma-treated water solutions PAW 4min, PAW 6min, and PAW 8min approximately 90 \%, 60 \%, and 40 \% germination respectively.\\ 
The length of the root and shoot were measured after two days of soaking seeds. Nearly 8 to 10 germinated seeds were taken to measure the root and shoot length on the third day (nearly 72 hours) of soaking seeds. After measuring the length of germinated grown seeds, we have taken an average of all these measured lengths and plotted data in Fig.~\ref{fig:fig6} for different plasma-treated water samples. We observed that seeds grown in PAW2 min have a larger (average) root and shoot length compared to other plasma-activated water samples. 
The root and shoot length (both) of seeds soaked into PAW 6min and PAW 8min are very small compared to seeds soaked in normal water and PAW 2min or PAW 4min solution. We can also see the growth of the root and shoot of the plant (crop) on different days (hours) in images shown in Fig.~\ref{fig:fig5}. We observe the maximum growth of plants in water that was treated for 2 min and 4 min and the minimum for 8 min treated water sample. It clearly indicates the negative effect of plasma-activated water on seed germination as well as on plant growth if it is treated for a longer time.\\
\begin{figure*} 
\centering
 \includegraphics[scale= 0.6600000]{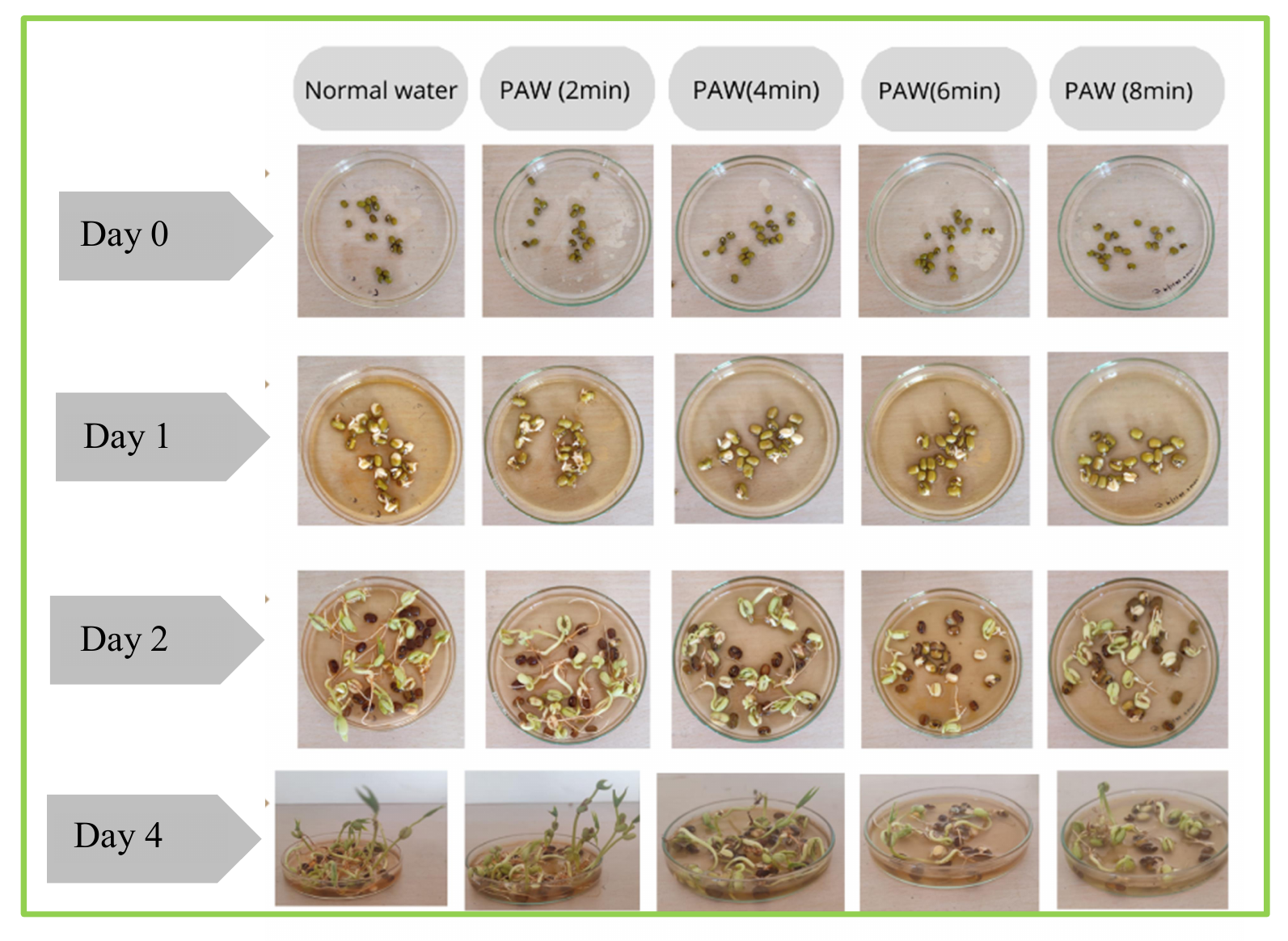}
\caption{\label{fig:fig7} Growth of Vigna radiata (moong seeds) in plasma-activated water samples}
\end{figure*} 
\begin{figure} 
\centering
 \includegraphics[scale= 0.3000000]{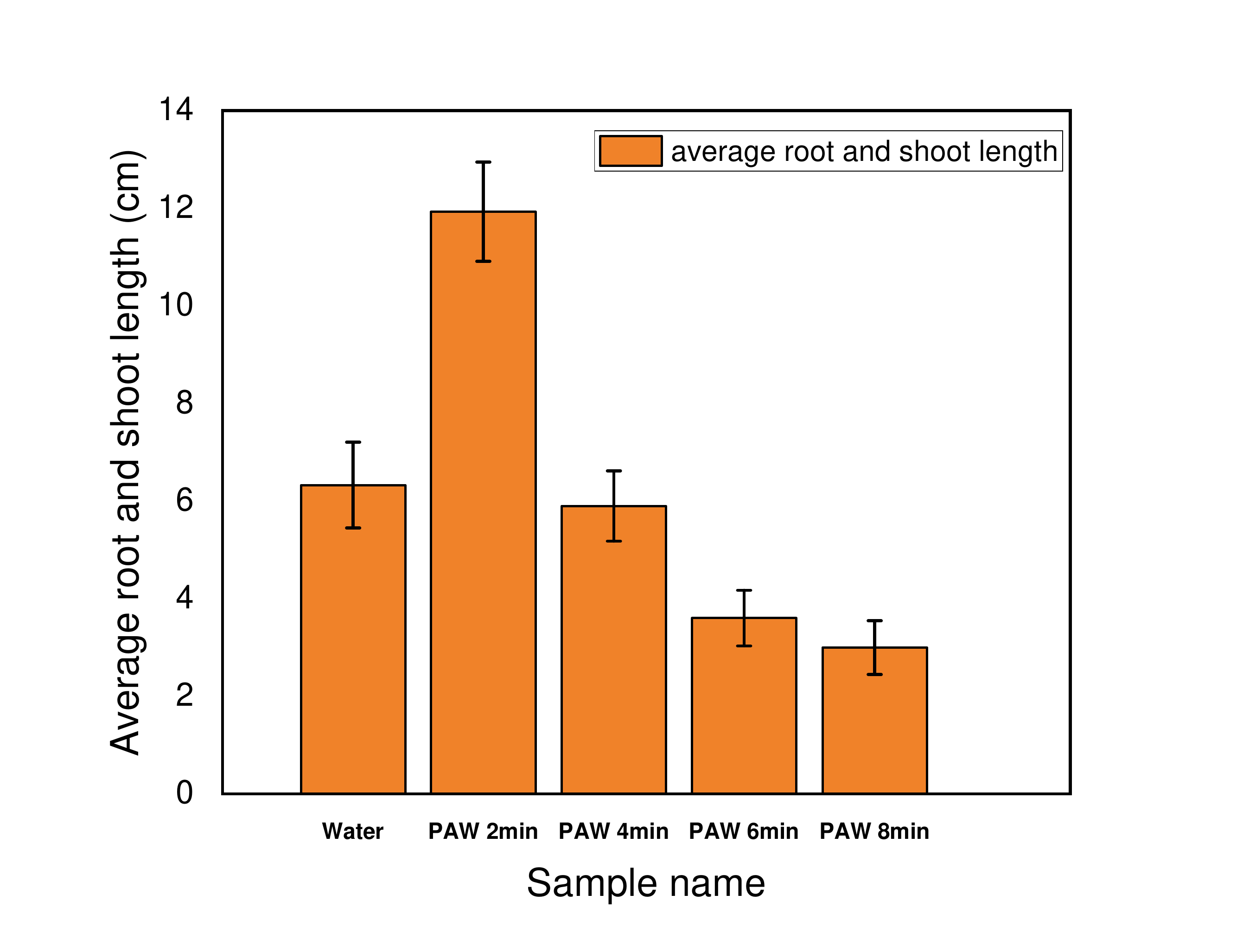}
\caption{\label{fig:fig8} Average root and shoot (both) length on the fourth day (after 94 hours) of soaking seeds in different plasma-activated water samples}
\end{figure} 
\begin{figure} 
\centering
 \includegraphics[scale= 0.29000000]{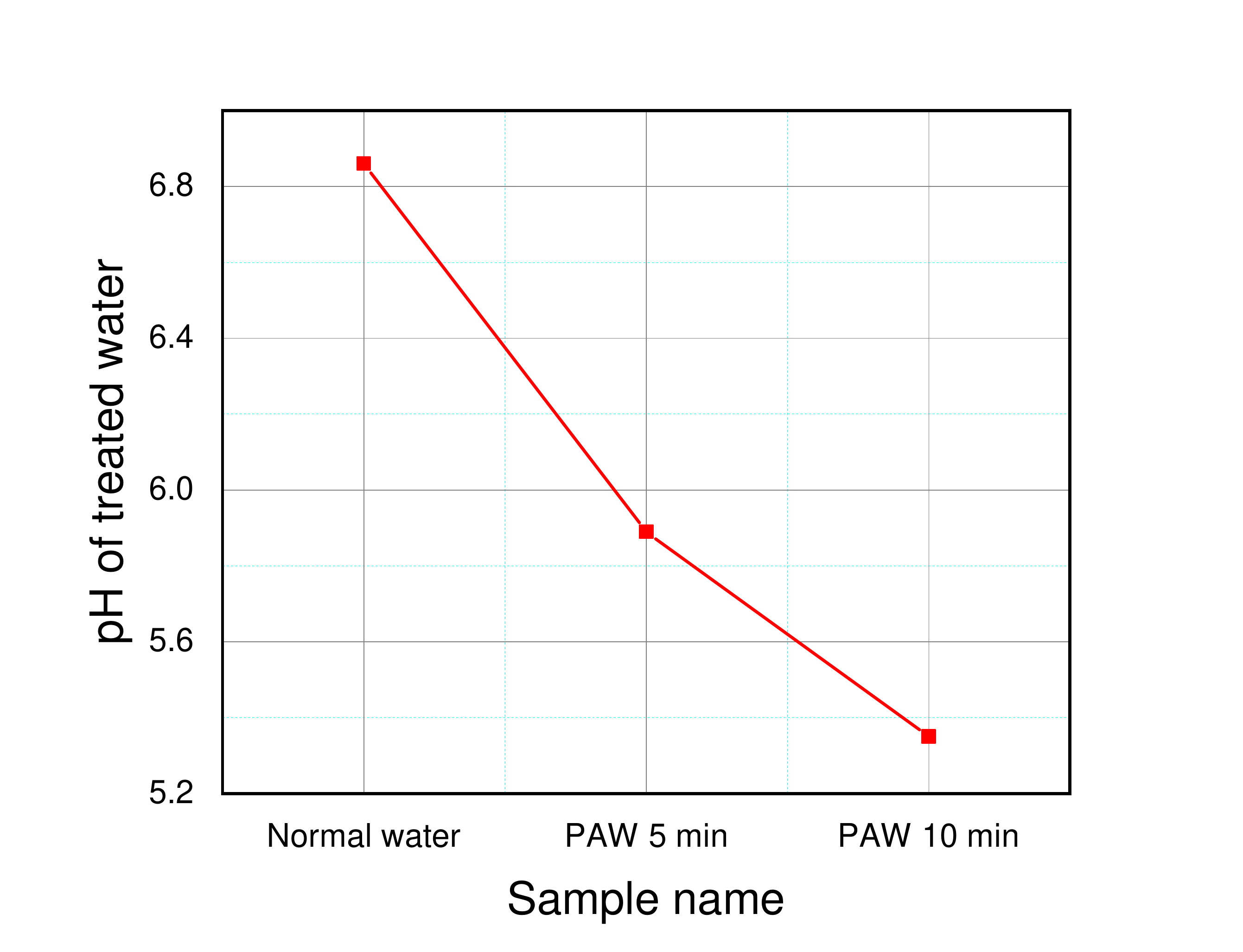}
\caption{\label{fig:fig9} Variation of pH of plasma activated water with time.}
\end{figure} 
\begin{table*}
\caption{Seeds germination percentage in untreated and plasma treated water}
\centering 
\begin{tabular}{|l|c|c|c|c|c}
\hline
\hline 
Sample &Germination ratio & Germination percentage& Germination ratio & Germination percentage \\
name & (Experiment - I)& (\%) & Experiment - II)& (\%) \\
\hline
Normal water& 19/20 & 95 & 20/20 & 100\\   
 PAW 2 min & 20/20& 100 & 20/20 & 100 \\
  PAW 4 min& 19/20&  95 & 18/20& 90 \\   
  PAW 6 min & 14/20& 70 &12/20 & 60 \\
  PAW 8 & 8/20& 40 & 9/20 & 45 \\
\hline \hline
\end{tabular}
\end{table*}
We performed another set of experiments with the same volume of water, same duty cycle, and same treatment time to explore the effect of the surrounding environment (temperature and humidity). This set of experiments was performed 15 days later when there was a change in the surrounding environment. We reported a change in temperature by 6 to 8 $^o$C between the first and second sets of experiments. The results obtained on seed germination and plant growth in this experiment were slightly different than those previously obtained results. The seed germination progress and plant growth in different plasma-activated water samples with time are depicted in Figure \ref{fig:fig7}. We see an effect of plasma-activated water on the growth of plants (crop) in the images of this figure. The average root and shoot (both) lengths were measured on the fourth day (96 hours) of seed soaking in the plasma-treated water samples. The average root and shoot length data are plotted in Figure \ref{fig:fig8}. The maximum growth of root and shoot was observed in PAW 2min and PAW 4min sample and minimum in PAW 8min.\\
As we have discussed the role of moisture (humidity) and temperature of the surrounding environment on the seed germination rate. The Difference in surrounding air temperature can also change the pH of water and concentrations of dissolved reactive species, therefore we expect slightly different root and shoot lengths in plasma-activated water in the second experiment. The pH of plasma-treated water decreases with increasing the treatment time as shown in Fig.\ref{fig:fig9}. We expect the role of nitrogen and oxygen compounds dissolved in the plasma-treated water to control the seed's germination rate and growth of plants. The amount of nitrogen compound is expected to increase with higher treatment time which reflects in decreasing pH value of the solution. 
\section{Conclusion and Future prospective} \label{sec:secVIII}
It is possible to summarize the main findings of this project by keeping the previous study and this primary study on plasma agriculture in mind. These points are as follows-
\begin{itemize}
    \item Low-temperature plasma technology has the potential to increase the productivity of crops instead of using any synthetic chemical fertilizers. 
    \item It is possible to fix air nitrogen by using plasma sources (gas discharges) and can dissolve the nitrogen compounds into tap water to make plasma-activated liquid fertilizer.
    \item The low-temperature plasma can also be used as a source of visible light, UV radiation, and pesticides for the growth and development of plants.
    \item The plasma-activated water can be used to increase the seed germination rate and percentages. 
    \item The chemical composition of plasma-activated water strongly depends on the surrounding environment. 
\end{itemize}
We discussed that plasma-activated water contains different reactive nitrogen and oxygen species. The $H_2O_2$ is known as a signal molecule in plant cells and plays a significant role in seed germination. It also regulates plant growth and development through various chemical reactions. Proper concentration of $H_2O_2$ causes softened the seed coat and allows the seed to absorb more oxygen. This results in increased seed germination speed. A higher concentration of $H_2O_2$ may be one of the causes of low seed germination in the present study at larger treatment times (PAW6 min and PAW8 min). Ozone ($O_3$) works as a disinfectant and reduces cellular toxification. Its concentration also regulates the growth of plants. We know that $NO_3^{-}$ is one of the absorbable forms of nitrogen for plants/crops. Nitric oxide is also responsible to promote seeds germination if its concentration is not exceeded to a proper concentration in plasma-treated water \cite{plantncert,plasmaagriculture1,biologybook1}. The concentration of $H_2O_2$, $NO_3^{-}$, $NH_3$, $O_3$, etc. strongly depends on the plasma treatment time. Therefore, we expect a higher amount of these compounds in a longer treated plasma-activated water (PAW 6 min or PAW 8 min).\\
We also noticed the change in pH of plasma treated water with changing temperature of tap water. It could be due to a change in the ionization process with increasing the solution temperature. The seed germination is also strongly affected by the acidic behavior (low pH). Therefore, the germination ratio is decreased with the increment of plasma interaction time. Hence we are getting less seed germination rate and plant growth (root and shoot length) in plasma-treated water for 6 or 8 min. \\
Plasma-treated water is considered as N-content rich liquid organic fertilizer \cite{plasmaagriculture1,plasmaagriculturereview1,plasmaagriculturereview2}.
It has been discussed that fertilizers are required to promote the growth and development of plants/crops. Therefore, we observed a higher growth rate (root and shoot) in plasma-activated water (PAW 2min and PAW 4min) than in normal water. The higher amount of N-content fertilizer (heavy dose) always affects the growth of plants/crops. A preferable amount of fertilizer is good to stimulate the growth-affecting factors of plants/crops, therefore, we observed lower plant growth in PAW 8min than in PAW 2min in the present work. We did not see a constancy in the observed results performed on different time intervals (gap of one week). There was a 15 days gap between performing different sets of experiments and the surrounding weather was getting changed every week. Therefore, we expect different chemical compositions of plasma-treated water by using the same atmospheric air plasma source while there is a change in the surrounding environment.\\
The primary experimental findings on plasma agriculture prove the great potential of plasma technology in the agriculture sector at every step from seed treatment to fruit/vegetable storage. However, there is a gap between the laboratory findings and their implementation in the field. There should be a bridge between the scientists and farmers to implement the plasma technology from the lab to the farm. To make plasma technology cost-effective and reliable for farmers, we must work on some important projects like designing and developing an appropriate plasma source to prepare plasma-activated water, need to make plasma-activated water a complete liquid fertilizer by adding required nutrients, need to operate the plasma sources by solar cells to reduce the cost of technology, need to prepare common data sets to use the such technique at every part of the globe, need a wide spectrum of research on the further development of technology, etc. In the future, we would be working on a few such projects with the specific objective to implement low-temperature plasma technology in the agriculture and food sector. 
\section{Acknowledgement} 
The authors are very grateful to Dr. Gajendra Singh, Dr.Raviprakash Chandra, Dr. Roli Mishra and Mrs. Nikita, Mr. Nilesh Patel for their assistance in chemical analysis, providing the laboratory facilities and fruitful discussion during the experiments at the Institute of Advanced Research, Gandhinagar, India.
\bibliography{Bibliography}
\end{document}